\documentclass[]{spie}  

\newcommand{\xx}[1]{ \textcolor{blue}{#1} }

\usepackage{amsmath,amsfonts,amssymb}
\usepackage{graphicx}
\usepackage[colorlinks=true, allcolors=blue]{hyperref}
\newcommand{\bftab}{\fontseries{b}\selectfont}
\usepackage{rotating}
\usepackage{mwe}
\usepackage{graphbox}
\usepackage{multirow}
\usepackage{authblk}
\usepackage{soul}
\usepackage{float}
\title{CATS v2: Hybrid encoders for robust medical segmentation}

\author[a]{Hao Li}
\author[b]{Han Liu}
\author[a]{Dewei Hu}
\author[b]{Xing Yao}
\author[b]{Jiacheng Wang}
\author[a,b]{Ipek Oguz}
\affil[a]{Department of Electrical and Computer Engineering, Vanderbilt University} 
\affil[b]{Department of Computer Science, Vanderbilt University}

\authorinfo{Further author information: (Send correspondence to Hao Li)\\Hao Li: E-mail: hao.li.1@vanderbilt.edu}

\begin{document} 
\maketitle

\begin{abstract}
Convolutional Neural Networks (CNNs)  exhibit strong performance in medical image segmentation tasks by capturing high-level (local) information, such as edges and textures. However, due to the limited field of view of convolution kernels, it is hard for CNNs to fully represent global information. Recently, transformers have shown good performance for medical image segmentation due to their ability to better model long-range dependencies. Nevertheless, transformers struggle to capture high-level spatial features as effectively as CNNs. A good segmentation model should learn a better representation from local and global features to be both precise and semantically accurate. In our previous work, we proposed CATS, which is a U-shaped segmentation network augmented with transformer encoder. In this work, we further extend this model and propose CATS \textit{v2} with hybrid encoders. Specifically, hybrid encoders consist of a CNN-based encoder path paralleled to a transformer path with a shifted window, which  better leverage both local and global information to produce robust 3D medical image segmentation. We fuse the information from the convolutional encoder and the transformer at the skip connections of different resolutions to form the final segmentation. The proposed method is evaluated on three public challenge datasets: Beyond the Cranial Vault (BTCV), Cross-Modality Domain Adaptation (CrossMoDA) and task 5 of Medical Segmentation Decathlon (MSD-5), to segment abdominal organs, vestibular schwannoma (VS) and prostate, respectively. Compared with the state-of-the-art methods, our approach demonstrates superior performance in terms of higher Dice scores. Our code is publicly available at \url{https://github.com/MedICL-VU/CATS}.
\end{abstract}


\section{introduction}
In recent years, deep learning (DL) has shown excellent performance in many medical image segmentation tasks \cite{liu2023medical}. As a fundamental unit of DL, convolutional neural networks (CNNs) are widely used for segmentation due to their ability to learn complex patterns and structures from medical datasets. By hierarchically learning parameters using both linear and non-linear layers, CNNs leverage both local and global information from images to predict segmentations. For instance, U-Net \cite{ronneberger2015u} is a popular architecture specifically designed for biomedical image segmentation. This U-shaped network consists of an encoder and a decoder, interconnected by skip connections. These connections ensure that high-resolution features are combined with upsampled low-resolution features to facilitate precise segmentation. Furthermore, variants of U-Net have demonstrated state-of-the-art performance across various medical image segmentation tasks and different imaging modalities \cite{zhang2021segmentation,li2021mri,li2021longitudinal,hu2021life,li2022human,li2022self,liu2022moddrop++}. However, due to the local receptive field of convolution kernels, convolutional encoders have limitations in modeling long-range dependencies and potentially missing out on global context in medical images.

Inspired by the success of the Vision Transformer (ViT) \cite{dosovitskiy2020image}, transformers have recently been adapted to the medical imaging field to produce high-quality segmentation \cite{shamshad2022transformers,chen2021transunet,hatamizadeh2022unetr}. These transformer-based methods process an input image/patch as a sequence of subpatches, rather than analyzing the entire input at once. With this property, the primary advantage of transformers is their ability to model long-range dependencies using the self-attention mechanism and to interact with all pixels in the image, in contrast to CNNs which possess a localized field of view. This global perspective is especially valuable in medical image segmentation, where contextual information from distant parts of the image can be important. However, ViT is computationally intensive and struggles to capture local information, especially for high-resolution medical data. As a variant of the ViT, the Swin Transformer \cite{liu2021swin} has shown good performance by computing representations hierarchically within shifted windows instead of applying self-attention to the entire image. Compared to ViT, the Swin Transformer reduces computational redundancies using the shifted window scheme, and it has been utilized in medical applications to produce robust segmentations from high-resolution medical data \cite{hatamizadeh2022swin,peiris2022robust,cao2022swin,zhou2023nnformer}. In addition to preserving global information, the shifted window approach also enhances the capture of local details. Given the importance of precise segmentation of anatomical structures and pathological regions in medical imaging, the ability to focus on fine-grained details is particularly advantageous for tasks like tumor and multi-organ segmentation \cite{hatamizadeh2022swin,peiris2022robust,cao2022swin, zhou2023nnformer}. 

Although the shifted window approach is effective, it may still not match the local specificity of a carefully designed CNN for certain medical image segmentation tasks, because fine-level details can be of paramount importance in medical imaging.
Thus, a hybrid approach that combines the strengths of both CNN and transformer might provide an optimal solution \cite{zhang2021transfuse,li2022cats,li2023promise}. This raises the question: could hybrid encoders incorporating Swin Transformers enhance the current segmentation networks used for 3D medical image segmentation?

In this work, we introduce a 3D segmentation network with hybrid encoders named CATS \textit{v2}. This is an improved version of our previous work, CATS (complementary \underline{C}NN \underline{a}nd \underline{t}ransformer encoders for \underline{s}egmentation) \cite{li2022cats}, and offers better performance. In particular, we replace the ViT with the Swin Transformer, which is used as an additional independent encoder in a U-shaped CNN. The multi-scale features extracted from the Swin Transformer are fused with the features from the CNN and then delivered to the CNN-based decoder for segmentation. We evaluate the proposed methods on three different segmentation tasks, including abdominal organs, vestibular schwannoma (VS), and prostate, where large inter-subject variations are present. We compare our model to state-of-the-art models on \xx{three} public datasets. The better performance of the proposed method in terms of Dice scores indicates that Swin Transformer improves the segmentation ability of existing segmentation networks with hybrid encoders. Moreover, our method has the potential to serve as a backbone for recent methods\cite{li2023promise,li2023assessing,yao2023false,wang2023novel,zhang2024segment}  based on the Segment Anything Model (SAM \cite{kirillov2023segment}) in the field of medical image segmentation.

\section{methods}

\subsection{Framework overview}
\begin{figure}[h]
\centering
\includegraphics[width=1\linewidth]{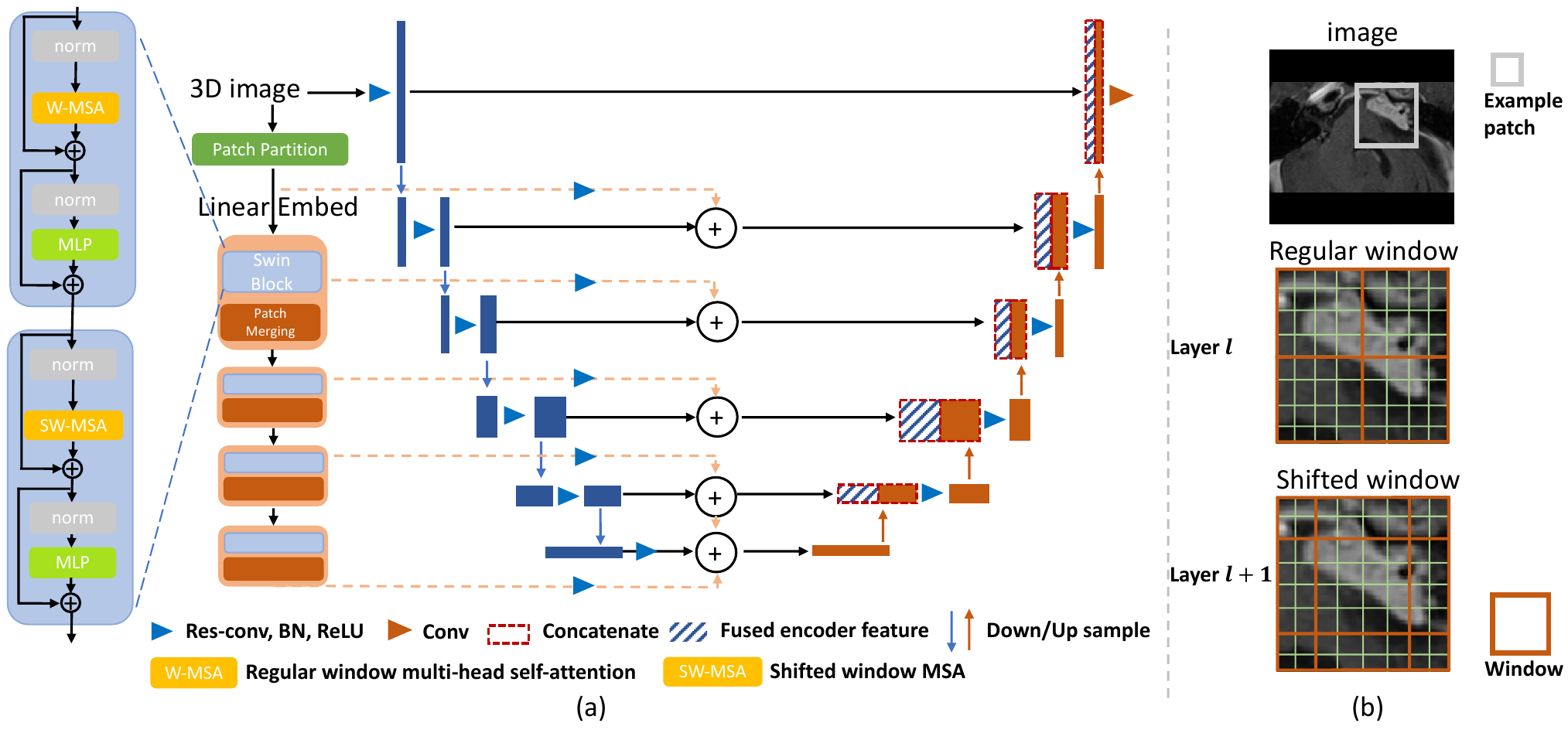}
\caption{\textbf{(a)} Proposed network architecture. \textbf{(b)} 2D illustrations of shifted window where self-attention is only computed within each non-overlapping local window. Note that the patch sizes vary.}
\label{network}
\end{figure}

Fig.~\ref{network} (a) shows the proposed segmentation network with hybrid encoders. Our model consists of two encoder paths: a CNN path and a transformer path with shifted window. The CNN-based encoder progressively encodes information using convolution and downsampling operations. On the Transformer path, the input images pass through the patch partition layer to reduce the dimension and visualize high-level features by a convolution operation and are then fed into the transformer blocks. The information from both paths is  fused at each level using addition operations, and this combined information is delivered to the CNN-based decoder to predict the final segmentation.

\subsection{Swin Transformer encoder}
The proposed Swin Transformer encoder is adopted from \cite{liu2021swin,hatamizadeh2022swin}. Specifically, the input of the Swin Transformer encoder is a 3D image, and a patch partition layer is applied to create a sequence of 3D patches/tokens with a given patch size. However, unlike ViT that flattens these patches and feeds them directly into the Transformer, non-overlapping local windows are created for efficient patch interaction modeling. Each local window goes through a linear projection layer to transform it into a sequence of token vectors. The transformed vectors are then processed by the self-attention mechanism of the transformer. Our encoder has four Swin blocks and each contains two successive transformer layers, i.e., regular window multi-head self-attention (W-MSA) and shifted window MSA (SW-MSA), which are shown in Fig.~\ref{network} (a). 

Fig.~\ref{network} (b) demonstrates the shifted window scheme for subsequent transformer layers. In the layer $l$ (W-MSA), we evenly partition the patch into subregions with same window size at each dimension. In the subsequent layer, $l+1$, the partitioned window regions are shifted by half of window size. The position of the windows is shifted to allow the model to gradually increase its receptive field and incorporate a more global context into its representations. To preserve the hierarchical structure of the encoder, a patch merging layer is employed at the end of each stage. This reduces the resolution of feature representations by a factor of 2, thereby decreasing the complexity and increasing the efficiency of the model. Following Hatamizadeh et al.\ \cite{hatamizadeh2022swin}, the embedding layer reduces the dimension of its input by half. Note that the linear projection layer enables the model to efficiently handle high-resolution inputs by reducing the dimensionality.

\subsection{Convolutional neural network architecture}
Fig.~\ref{network} (a) also shows the proposed CNN, which is adapted from the 3D U-Net and its variants \cite{cciccek20163d,li2021unsupervised}. Max-pooling and deconvolution operations are employed for down-sample and up-sample, respectively. The feature maps from the highest level are sent directly to the decoder, while feature maps from the lower levels are combined with encoded information from the Swin Transformer encoder path via addition. This fused information is then delivered to the decoder using skip connections, following the pattern of the 3D U-Net \cite{cciccek20163d} to produce the final segmentation.

\subsection{Datasets}

We use three publicly available datasets in our experiments. 

\begin{itemize}
\item The \textbf{BTCV} \cite{landman2015miccai} dataset contains
30/20 subjects with abdominal CT images for training/testing,
with 13 different organs labeled by experts. The results are obtained from the official leaderboard.

\item \textbf{CrossModa} \cite{dorent2023crossmoda}
has 105 contrast-enhanced T1-weighted MRIs with manual labels for vestibular schwannomas (VS). We split the dataset into 55/20/30 for training/validation/testing. 

\item \textbf{MSD-5} \cite{antonelli2022medical} consists of 32 MRIs with manual prostate labels. 2 MRIs in validation were excluded due to the wrong labels being provided in the public dataset. We use this dataset in a 5-fold cross-validation framework, and follow the setting in nnUnet\cite{isensee2019automated}. 

\end{itemize}

Dice score, average surface distance (ASD) and 95-percent Hausdorff distance (HD95) are used as evaluation metrics. The details of preprocessing steps for all datasets can be found in the original CATS paper \cite{li2022cats}.

\subsection{Implementation details} We followed the implementation settings in CATS \cite{li2022cats} for our experiments for a fair comparison. Briefly, we normalized the image intensity to range [0, 1]. The constant learning rate was set to 0.0001. Training batch size was 2 for all experiments which are conducted on Pytorch, MONAI and an NVidia Titan RTX GPU.
\section{RESULTS}

\subsection{BTCV results} 
\begin{figure}[h]
\centering
\includegraphics[width=.96\linewidth]{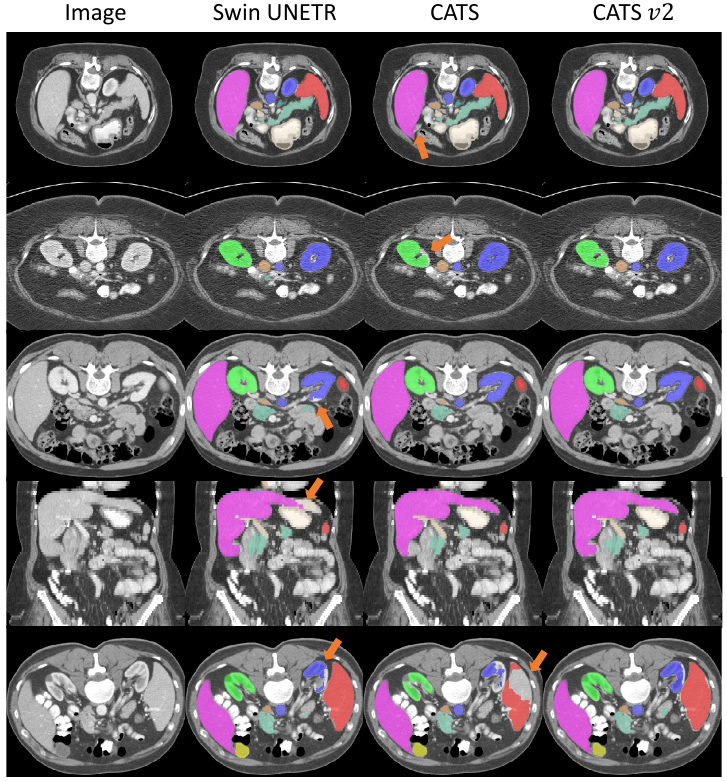}
\caption{Qualitative results in BTCV. Some major differences are highlighted by orange arrows.}
\label{btcv_fig}
\end{figure}

The quantitative and qualitative results of BTCV dataset are shown in Tab.~\ref{btcv} and Fig.~\ref{btcv_fig}, respectively. The compared methods include TransUNet \cite{chen2021transunet}, UNETR \cite{hatamizadeh2022unetr}, Swin UNETR \cite{hatamizadeh2022swin}, CATS \cite{li2022cats}, and the proposed CATS \textit{v2}. Briefly, UNETR \cite{hatamizadeh2022unetr} is composed of a ViT encoder and a CNN decoder, while Swin UNETR replaces the ViT with a Swin encoder. Similarly, CATS \cite{li2022cats} is built upon the 3D U-Net \cite{cciccek20163d} and integrates a ViT encoder. The proposed CATS \textit{v2} employs a Swin encoder as the upgrade. 

From Tab.~\ref{btcv}, the proposed CATS \textit{v2} achieves the best overall performance among the state-of-the-art compared methods (the `Avg.' column). In the comparison between Swin UNETR and proposed CATS \textit{v2}, we observe the improvements in 8 out of 13 organs when a CNN encoder is integrated. Furthermore, the proposed CATS \textit{v2} outperforms original CATS in 7 out of 13 organs, with  larger improvements observed in organs of smaller volume, such as the gallbladder, and the right and left adrenal glands. These improvements suggest that the Swin encoder could further refine the local details. Fig.~\ref{btcv_fig} shows qualitative results, with major differences highlighted by orange arrows. Compared to the Swin UNETR and the original CATS, our proposed model produces smoother results.

\begin{table*}[t]
\caption{Mean Dice scores in BTCV dataset. Bold numbers denote the highest Dice scores. The results of TransUNet are directly copied from \cite{chen2021transunet}. The experiments follow the public pipeline of Swin UNETR\cite{hatamizadeh2022swin}. The organs from left to right are: spleen, right and left kidney, gallbladder, esophagus, liver, stomach, aorta, inferior vena cava, portal vein and splenic vein, pancreas, right and left adrenal gland, and overall average. Bold numbers indicate the best performance. The results can be found on the official leaderboard.}
\label{btcv}
\begin{center}
    \begin{tabular}{  l  c l l l l l l l }
    \hline
    \hline
     Method & & Spl & RKid & LKid & Gall & Eso & Liv & Sto\\
     \hline
     
     TransUNet \cite{chen2021transunet}& \vline& 85.1 & 77.0 & 81.9 & 63.1 & - & 94.1 & 75.6 \\
     
    UNETR \cite{hatamizadeh2022unetr}& \vline& 93.4 & 85.5 & 87.6 & 61.9 & 74.7 & 95.7 & 76.8 \\

    Swin UNETR \cite{hatamizadeh2022swin}& \vline& \textbf{95.9} & 87.8 & 92.9 & 65.7 & 77.2 & 96.5 & 83.3 \\
     
    CATS & \vline& 95.8 & \textbf{90.2} & \textbf{93.4} & 65.9 & 77.1 & \textbf{96.8} & 83.0\\

    CATS \textit{v2} &\vline & 94.8 & 87.1 & 93.2 & \textbf{70.7} & \textbf{78.1} & 96.7 & \textbf{85.8}\\

  \end{tabular}

      \begin{tabular}{  l  c l l l l l l l}
\hline

      & & Aor & IVC & Veins &Pan &RAG &LAG&Avg.\\
     \hline
     
     TransUNet \cite{chen2021transunet}& \vline& 87.2 & - & - 
     & 55.9 & - & - & 77.5 \\
     
    UNETR \cite{hatamizadeh2022unetr}& \vline&  85.2 & 77.2 &69.8 & 61.5 & 64.4 & 59.4 &76.9 \\

    Swin UNETR \cite{hatamizadeh2022swin}& \vline&  85.5 & 82.8 & 75.1 & 72.5 & \textbf{74.0} & \textbf{72.0} & 81.6 \\
     
    CATS & \vline& \textbf{88.6} & \textbf{83.1} & 76.9 & 73.8 & 70.2 & 62.6 & 81.4\\

    CATS \textit{v2} &\vline & 88.0 & 82.5 & \textbf{77.0} & \textbf{76.1} & 72.2 & 66.3 &  \textbf{82.2} \\

    \hline

  \end{tabular}

\end{center}
\end{table*}
\begin{table}[H]
\caption{Quantitative results in CrossMoDA dataset, presented as $mean (std.dev.)$. Bold numbers indicate the best performance.}
\label{moda}
\begin{center}
    \begin{tabular}{  l  c c c c}
    \hline
    \hline
     Method & & Dice & ASD & HD95\\
     \hline
     
     2.5D CNN \cite{shapey2019artificial} & \vline& 0.856 (1.000) & 0.69 (1.20) & 3.5 (5.2) \\
     
     TransUNet \cite{chen2021transunet} & \vline& 0.792 (0.234) & 7.86 (27.6) & 12 (31) \\
     
     UNETR \cite{hatamizadeh2022unetr}& \vline& 0.772 (0.139) & 7.95 (14.2) & 26 (43) \\
     
     CATS \cite{li2022cats} & \vline& 0.873 (0.088) & \textbf{0.48 (0.63)} & 2.6 \textbf{(3.6)} \\

      CATS \textit{v2} & \vline& \textbf{0.886 (0.076)} & \textbf{0.48} (0.79) & \textbf{2.4} (4.0) \\          
    \hline

  \end{tabular}
\end{center}
\end{table}

\subsection{CrossMoDA results}

\begin{figure}[t]
\centering
\includegraphics[width=0.96\linewidth]{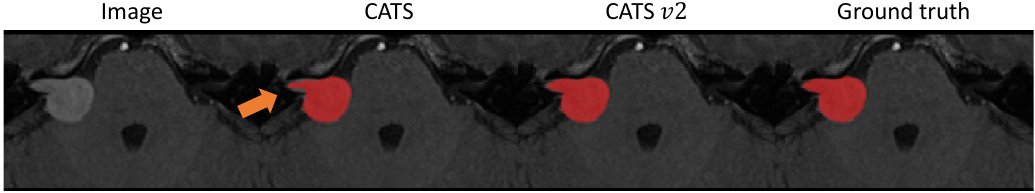}
\caption{Qualitative results in CrossMoDA. Local segmentation errors are highlighted with arrows.}
\label{moda_fig}
\end{figure}

The quantitative results for the CrossMoDA dataset are presented in Tab.~\ref{moda}. We compare the models against a 2.5D CNN model \cite{shapey2019artificial}, which was specifically designed to segment VS from MRIs characterized by substantial discrepancies between in-plane resolution and slice thickness, which is a common feature of this dataset. We observe that this CNN-only network  performs better than the transformer-based encoders \cite{chen2021transunet,hatamizadeh2022unetr} for this task. The original CATS \cite{li2022cats} model outperformed the 2.5D CNN. With subsequent enhancements, our updated CATS \textit{v2} model further refined the quality of segmentation, delivering the highest performance in terms of Dice score. Fig.~\ref{moda_fig} shows the qualitative results of VS segmentation. While the original CATS model undersegments the VS (marked by arrow), the proposed CATS \textit{v2} effectively compensates for this limitation and produces robust results that align more closely with the ground truth segmentations.

\subsection{MSD-5 results}
\begin{figure}[b]
\centering
\includegraphics[width=0.96\linewidth]{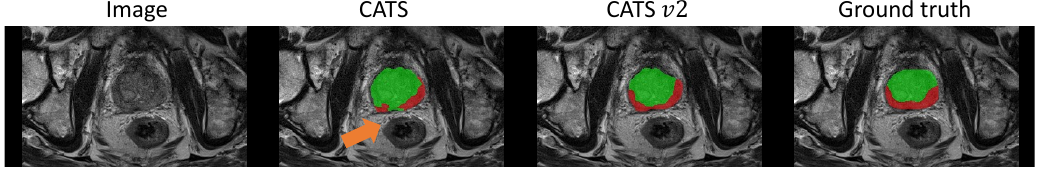}
\caption{Qualitative results in MSD-5. Local segmentation errors are highlighted with arrows. Red and green labels denote  the peripheral zone (PZ) and the transition zone (TZ), respectively.}
\label{msd_fig}
\end{figure}

We compared the nnUnet \cite{isensee2019automated}, TransFuse \cite{zhang2021transfuse} and CATS \cite{li2022cats} to our proposed method for the prostate segmentation task in Tab.~\ref{prostate}. CATS \textit{v2} has the highest Dice scores on all labels, i.e., both the peripheral zone (PZ) and  the transition zone (TZ). This dataset was chosen because of the inherent challenge in segmenting two closely adjoined regions that exhibit considerable inter-subject variability. The qualitative improvements between original CATS and CATS \textit{v2} are shown in Fig.~\ref{msd_fig}. A more robust segmentation is produced by the proposed method by correcting the false positives.

\begin{table}[b]
\caption{Mean Dice scores in MSD-5 dataset. PZ and TZ denote the peripheral zone and the transition zone, respectively. Bold numbers indicate the best performance.}
\label{prostate}
\begin{center}
    \begin{tabular}{  l  c c c c}
    \hline
    \hline
     Method & & PZ & TZ & Avg.\\
     \hline
     2D nnUnet \cite{isensee2019automated} & \vline& 0.6285 & 0.8380 & 0.7333 \\
     
     3D nnUnet \cite{isensee2019automated}& \vline& 0.6663 & 0.8410 & 0.7537 \\
     
     TransFuse \cite{zhang2021transfuse} & \vline& 0.6738 & 0.8539 & 0.7639 \\
     
     CATS \cite{li2022cats} & \vline& 0.7136 & 0.8618 & 0.7877 \\

     CATS \textit{v2} & \vline& \bftab0.7356 & \bftab0.8713 & \bftab0.8034 \\

    \hline

  \end{tabular}
\end{center}
\end{table}

\section{Discussion and conclusion}
In this work, we introduce CATS \textit{v2}, which is a segmentation network with hybrid encoders, specifically, a U-shaped CNN complemented with a Swin Transformer. We evaluated our proposed methods on three public datasets that present large inter-subject variations. Our proposed model outperforms state-of-the-art models on each task. Relative to the original CATS, the Swin Transformer is able to further enhance the segmentation ability of the encoder. However, we observe inconsistent improvements in the BTCV dataset, indicating that one encoder may dominate the results. Exploration of other fusion strategies to overcome this issue remains as future work. In addition, due to the use of hybrid encoders as well as deeper architecture design, our proposed network might require slightly more computational resources than the original CATS. In the future work, we aim to design a light-weight model for 3D medical image segmentation.

{\bf Acknowledgements.}
This work was supported, in part, by NIH grant U01-NS106845, NIH grant R01-NS094456, and NSF grant 2220401.

\bibliography{refs} 
\bibliographystyle{spiebib} 

\end{document}